# Polarization-Sensitive Module for Optical Coherence Tomography Instruments

Po-Yi Lee, Chuan-Bor Chueh, Milen Shishkov, Sebastián Ruiz-Lopera, Tai-Ang Wang, Hsiang-Chieh Lee, Teresa Chen, Brett E. Bouma, and Martin Villiger

***Abstract*—Polarization-sensitive optical coherence tomography (PS-OCT) extends OCT by analyzing the polarization states of backscattered light to quantify tissue birefringence. However, conventional implementations require polarization-diverse detection and are therefore incompatible with most commercial OCT systems. As a result, PS-OCT has largely remained restricted to specialized research groups, limiting its broader scientific and clinical use. Here, we present a modular PS-OCT framework that integrates with a standard spectral-domain OCT platform through a detachable rotating achromatic half-wave plate in the sample arm. This waveplate modulates both incident and reflected polarization states. Three or more repeated measurements at distinct waveplate orientations enable reconstruction of the sample's round-trip Jones matrix and the corresponding polarization properties. To mitigate random phase variations between repeated measurements, we introduce a phase optimization strategy. We validate the framework with imaging of birefringent phantoms and the human retina *in vivo*, demonstrating reliable reconstruction of retardance and optic axis orientation. In this study, we implemented a polarization-sensitive module on the SPECTRALIS platform with minimal hardware modification. The framework could potentially be extended to other OCT systems with appropriate mechanical integration and access to acquisition data. By reducing the hardware complexity typically associated with conventional PS-OCT, this framework may facilitate broader adoption of PS-OCT imaging in both research and clinical settings.**

## I. Introduction

Optical coherence tomography (OCT) is a non-invasive imaging technique that generates cross-sectional images of biological tissues at micrometer resolution. Since its introduction in 1991 [1], OCT has become an indispensable tool in biomedical imaging, with applications ranging from ophthalmology [2, 3] to cardiology [4, 5], gastrointestinal [6] and pulmonary [7] imaging, and dermatology [8]. Polarization-sensitive OCT (PS-OCT) extends OCT by analyzing the polarization states of backscattered light, enabling tomographic reconstruction of tissue polarization properties such as birefringence and optic axis orientation [9, 10]. These polarimetric measurements offer insights into tissue microstructure beyond conventional OCT resolution, revealing information about tissue integrity, composition, and biomechanical factors. PS-OCT has been used to investigate a variety of tissues, including skin and skin-related conditions such as burns and scars [11-13], coronary arteries [14, 15], pulmonary airways [16, 17], and ocular tissues in studies of glaucoma [18, 19], age-related macular degeneration [20], and myopia [21, 22]. Despite these advances, PS-OCT has remained largely confined to research settings. Current implementations rely on custom-built systems with multiple polarization states for illumination and/or polarization-diverse detection channels [10]. The conventional PS-OCT frameworks are fundamentally incompatible with the widely deployed OCT platforms used in clinical and translational research. This incompatibility has hindered the development of commercially viable, clinically adaptable PS-OCT solutions, slowing progress across multiple fields of medicine. To enable broad adoption, PS-OCT requires

 This work was supported in part by the U.S. National Institute of Health under Grants R01EB033321 and P41EB015903; the Taiwan National Science and Technology Council under Grant NSTC113-2917-I-002-019; Royal Palm Elite Doctoral Scholarship from National Taiwan University; and Heidelberg Engineering GmbH for in-kind support. (*Corresponding author*: Po-Yi Lee)

This work involved human subjects in its research. Approval of all ethical and experimental procedures and protocols was granted by the Mass General Brigham Institutional Review Board.

Po-Yi Lee, Milen Shishkov, and Martin Villiger are with Wellman Center for Photomedicine, Massachusetts General Hospital, Somerville, MA 02145 (email: plee26@mgh.harvard.edu; mshishkov@mgh.harvard.edu; mvilliger@mgh.harvard.edu).

Sebastián Ruiz-Lopera is with Wellman Center for Photomedicine, Massachusetts General Hospital, Somerville, MA 02145 and also with Electrical Engineering and Computer Science, Massachusetts Institute of Technology, Cambridge, MA 02139 (email: srulo@mit.edu).

Brett E. Bouma is with Wellman Center for Photomedicine, Massachusetts General Hospital, Somerville, MA 02145 and also with Institute for Medical Engineering and Science, Massachusetts Institute of Technology, Cambridge, MA 02139 (email: bouma@mgh.harvard.edu).

Chuan-Bor Chueh and Tai-Ang Wang were with Wellman Center for Photomedicine, Massachusetts General Hospital, Somerville, MA 02145. Both are now with Graduate Institute of Photonics and Optoelectronics, National Taiwan University, Taipei, Taiwan 10617 (email: f06941077@ntu.edu.tw; d10941014@ntu.edu.tw).

Hsiang-Chieh Lee is with Graduate Institute of Photonics and Optoelectronics and Department of Electrical Engineering, National Taiwan University, Taipei, Taiwan 10617 (email: hclee2@ntu.edu.tw).

Teresa Chen is with Department of Ophthalmology, Massachusetts Eye and Ear, Boston, MA 02114 (email: Teresa_Chen@meei.harvard.edu).

a technical transition paradigm that integrates seamlessly with existing OCT systems while providing advanced polarization contrast. In this study, we introduce RevoPol, a novel PS-OCT framework that enables commercial spectral-domain OCT instruments to perform seamless polarimetric measurements through modular hardware modifications. By expanding PS-OCT capabilities to standard OCT platforms, RevoPol facilitates rapid deployment across diverse biomedical applications, from ophthalmology to other tissue imaging domains, accelerating both research and clinical translation.

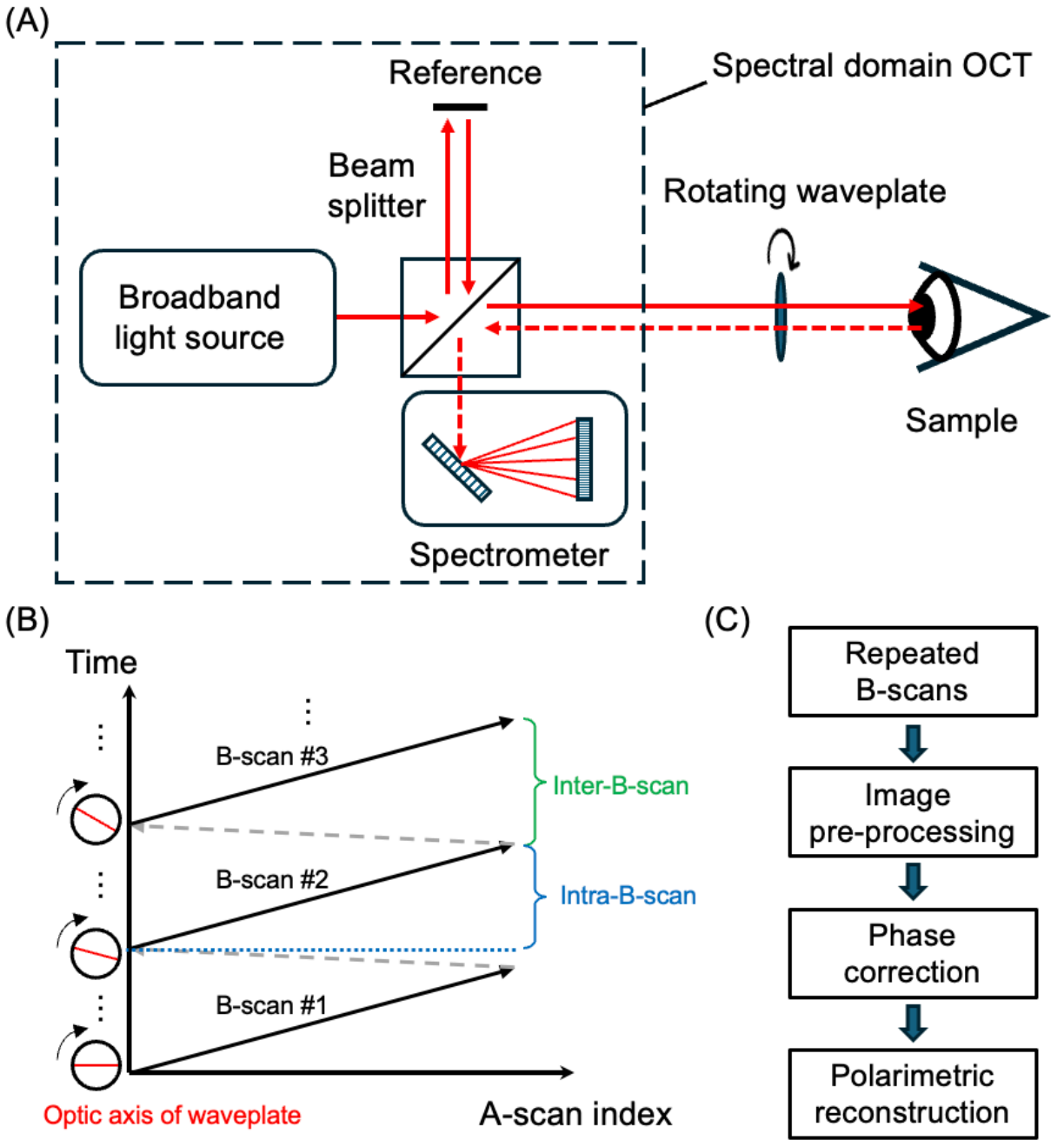

Fig. 1. Schematic of the RevoPol framework. (A) A rotating waveplate, placed in the sample arm, modulates the polarization state of the probing light and enables extraction of tissue polarization properties on a standard spectral-domain OCT platform. (B) In RevoPol, the waveplate rotates continuously throughout the imaging process. The rotation increment between matching A-lines of consecutive B-scans (inter-B-scan) is constant, where each scanning cycle includes data acquisition and scanner flyback (indicated by dashed arrows). Because the waveplate continues to rotate during the flyback interval, the effective rotation occurring within a single B-scan (intra-B-scan) is slightly smaller than the total inter-B-scan rotation. The reconstructed sample optic axis orientation is referenced to the initial waveplate rotation and requires compensation. (C) RevoPol first acquires repeated B-scans while the waveplate rotates. The acquired images then undergo image pre-processing, including image registration and coherency filtering, followed by phase correction and polarimetric reconstruction.

## II. Methods

### A. Imaging Framework and Setup

The novel PS-OCT framework, RevoPol, relies on inserting a rotating waveplate into the sample arm of a spectral-domain OCT instrument (**Fig. 1**). Specifically, employing an achromatic half-wave plate preserves the linearity of the source polarization state upon transmission while rotating it by twice the angle between the incident state and the waveplate's fast axis. During repeated cross-sectional imaging, the changing waveplate orientation modulates the polarization state of the illuminating and backscattered light to and from the sample. Like all OCT systems, the coherent measurement scheme is inherently polarization-sensitive: only the component of backscattered light matching the reference polarization contributes to the interference signal. In the RevoPol framework, scatterers in non-birefringent tissues exhibit identical interference signals across all waveplate orientations, while in birefringent tissues the interference signals vary with waveplate orientation. When birefringent and non-birefringent structures coexist at different depths, the depth-resolved OCT signal enables their distinction. By acquiring at least three repeated cross-sectional images at the same slow-scan location during waveplate rotation, the RevoPol framework reconstructs the tissue's polarization properties.

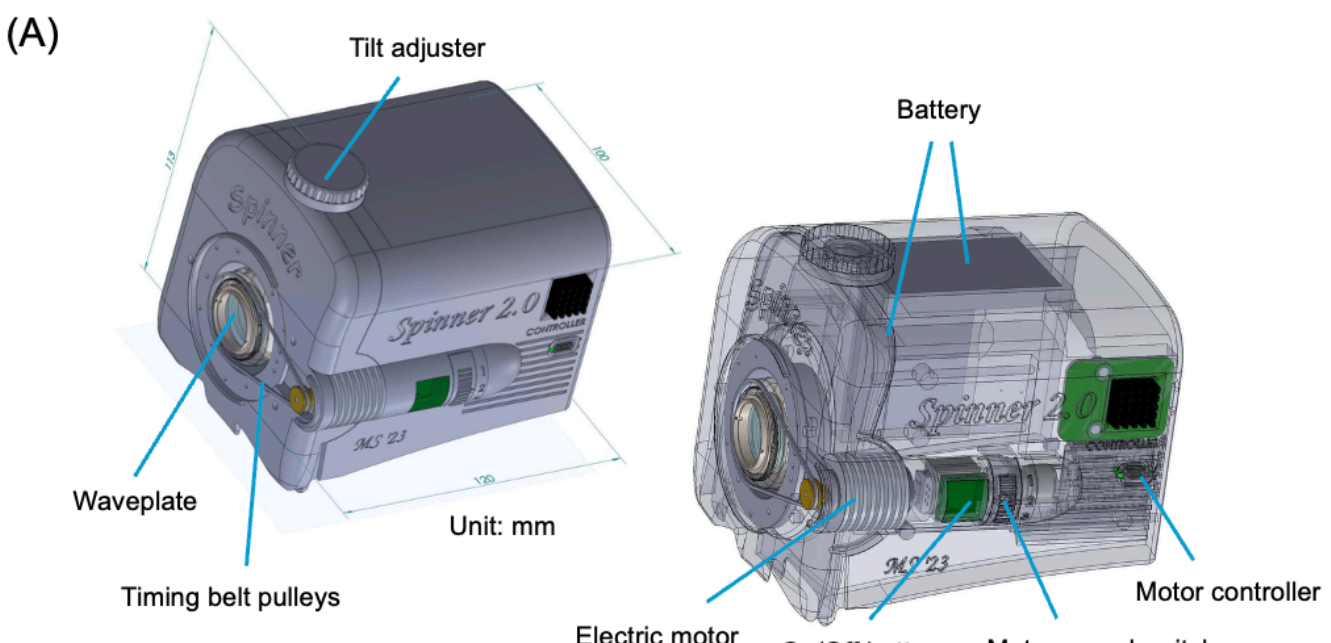

(B)

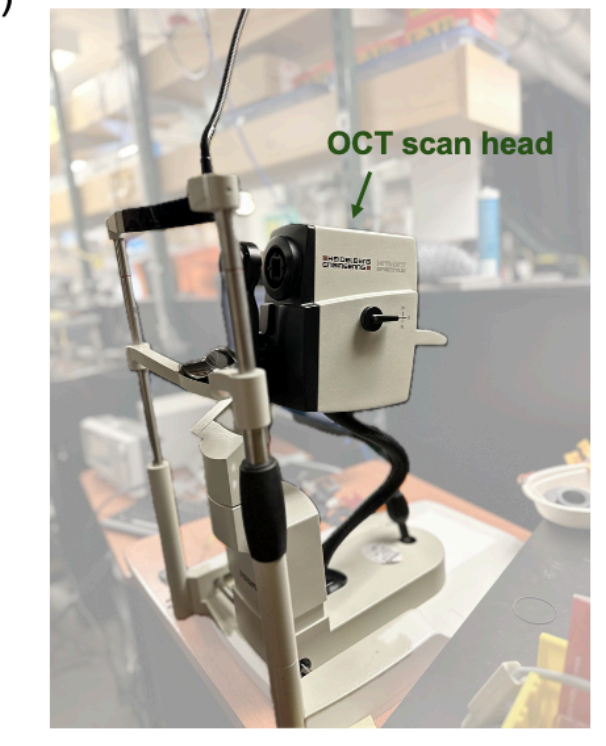

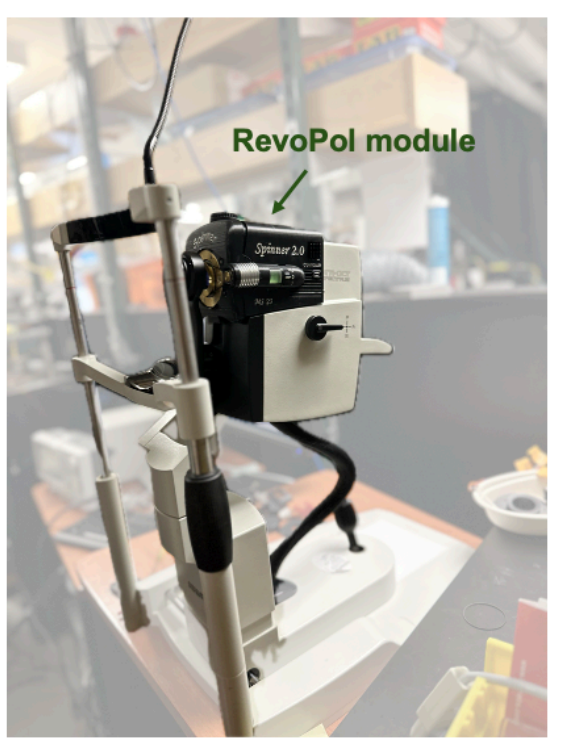

Fig 2. RevoPol module for the Spectralis HRA+OCT instrument. (A) The module consists of an achromatic half-wave plate driven by a belt-coupled DC motor. A motor controller, programmed via a micro-USB connection, allows the user to select rotation speeds using a dial on the module. The unit is powered by rechargeable batteries that can be recharged, eliminating the need for a wall outlet. The module operates independently and does not require synchronization with the OCT instrument. The tilt adjuster lowers the internal reflection from the waveplate. (B) A physical gap between the scan head and the objective lens facilitates integration of the module, which can be easily attached to and detached from the scan head.

To implement the rotating waveplate function in the RevoPol framework, we developed a module consisting of an achromatic half-wave plate (#48-500, Edmund Optics, USA) driven by a belt-coupled DC motor, with rotation speed controlled by a motor controller (**Fig. 2A**). The module was tailored for integration with the Spectralis HRA+OCT instrument (Heidelberg Engineering, Germany) by insertion into the free-space section of the sample arm (**Fig. 2B**). The Spectralis combines a spectrometer-based OCT system—operating at 85 kHz A-line rate with a source centered on 870 nm—with a confocal scanning laser ophthalmoscope (SLO) providing a 30° field-of-view and 30-Hz active eye tracking. Simultaneous SLO imaging enables real-time eye motion tracking, which stabilizes OCT imaging by dynamically correcting scan locations. This capability also supports OCT angiography (OCTA) acquisition, which uses repeated scans at each sampling location. Leveraging this OCTA mode, RevoPol continuously rotates the waveplate during scanning and reconstructs tissue polarization properties by acquiring at least three repeated cross-sectional images at the same location, each corresponding to a different waveplate orientation. The Spectralis evaluates the quality of each set of repeated measurements immediately after acquisition. If a set fails to meet quality control—for example, due to motion artifacts or signal dropout—it is discarded and reacquired. Using research software access authorized by Heidelberg Engineering, B-scan timestamps and processed OCT fringe data, including k-space interpolation and built-in OCT dispersion compensation, were exported for offline analysis. Dispersion introduced by the inserted waveplate was neglected, as it is achromatic by design and has a small thickness. The associated signal formalism and processing workflow are described in the following sections.

### B. Terminology

**A-line:** A depth-resolved signal profile at a single lateral position.

**B-scan**: A cross-sectional image acquired by sequentially acquiring A-lines while scanning laterally along the fast axis.

**Repeated measurements**: Multiple acquisitions at the same B-scan position, typically performed in OCTA mode while the waveplate rotates. A set of repeated measurements is used to reconstruct a polarization-resolved cross-sectional image.

**C-scan**: A three-dimensional dataset formed by acquiring multiple sets of repeated measurements along the slow axis. The index of the reconstructed cross-sectional images is denoted as the slow-axis number (slow-axis #).

**Surface retardance**: The cumulative retardance measured at the sample surface, introduced by birefringent tissues along the imaging path (e.g., cornea in retinal imaging).

### C. Jones Matrix Reconstruction

The RevoPol framework performs repeated measurements through a rotating waveplate placed in the sample arm, each using a single input and detection state. Measurements through distinct waveplate orientations effectively correspond to different polarization states incident on the sample, and the collected back-scattered light interferes with the reference light. Only the portion of the scattered light matching the polarization of the reference contributes to the interference signal, serving as a form of polarization detection. Algebraically, the recovered complex-valued tomogram can be expressed as

$$m(z,n) = \underbrace{\mathbf{f}^{\dagger} \cdot \mathbf{Q}^{\mathrm{T}}(\alpha + n\Delta\alpha)}_{\mathbf{f'}_n{}^{\dagger}} \cdot \mathbf{J}_S(z) \cdot \underbrace{\mathbf{Q}(\alpha + n\Delta\alpha) \cdot \mathbf{e}}_{\mathbf{e'}_n}, \tag{1}$$

where $\mathbf{Q}(\alpha + n\Delta\alpha)$ describes the transmission through the rotating half-wave plate at an initial angle $\alpha$ increasing by $\Delta\alpha$ between the sequential measurements, indexed by $n$. $\mathbf{e}$ is the complex Jones vector of the source polarization state, and $\mathbf{f}$ is the corresponding reference state. $\mathbf{e'}_n$ and $\mathbf{f'}_n$ denote the polarization states of the illumination and reference beams, respectively, after (reverse) transmission through the waveplate at the n$^{\mathrm{th}}$ measurements. $\mathbf{J}_S(z)$ represents the 2×2 round-trip Jones matrix of the tissue at depth $z$. Without loss of generality, the OCT system's free-space interferometer is linearly polarized and identical to the source polarization state (i.e., $\mathbf{e} = \mathbf{f}$).

Invoking the Kronecker product rule, each individual tomogram can be described as a linear projection of the unknown, vectorized Jones matrix $\overrightarrow{\mathbf{J}_S}(z)$ onto the known 4×1 sensing vector $\overrightarrow{\mathbf{g}_n}$, such that

$$m(z,n) = \underbrace{\mathbf{e'}_n{}^{\mathrm{T}} \otimes \mathbf{f'}_n{}^{\dagger}}_{\overrightarrow{\mathbf{g}_n}} \cdot \overrightarrow{\mathbf{J}_S}(z). \tag{2}$$

Here, the sensing vector $\overrightarrow{\mathbf{g}_n}$ encodes the illumination and detection polarization states at a specific waveplate orientation.

A sequence of tomographic measurements $\mathbf{m}$ forms a linear system that relates the unknown, vectorized Jones matrix $\overrightarrow{\mathbf{J}_S}(z)$ to the known sensing matrix $\mathbf{G}$:

$$\mathbf{m}(z) = \begin{bmatrix} m(z,1) \\ m(z,2) \\ \vdots \end{bmatrix} = \underbrace{\begin{bmatrix} \mathbf{e'}_1{}^{\mathrm{T}} \otimes \mathbf{f'}_1{}^{\dagger} \\ \mathbf{e'}_2{}^{\mathrm{T}} \otimes \mathbf{f'}_2{}^{\dagger} \\ \vdots \end{bmatrix}}_{\mathbf{G}} \cdot \overrightarrow{\mathbf{J}_S}(z) = \mathbf{G} \cdot \overrightarrow{\mathbf{J}_S}(z). \tag{3}$$

It is important to note that when the sensing matrix is constructed using only a rotating waveplate with fixed retardance, its maximum column rank is three, regardless of the number of waveplate orientations. The intrinsic round-trip symmetry of OCT further imposes that $\mathbf{J}_S(z) = \mathbf{J}_S{}^{\mathrm{T}}(z)$, and accordingly $j_{21} = j_{12}$ [23, 24]. Thus, instead of four independent elements, each vectorized Jones matrix reduces to three independent complex-valued scalars. This reduction can be expressed using a modified sensing matrix $\mathbf{G'}$ with only three columns:

$$\mathbf{m}(z) = \mathbf{G} \cdot \begin{bmatrix} j_{11}(z) \\ j_{21}(z) \\ j_{12}(z) \\ j_{22}(z) \end{bmatrix} = \underbrace{\mathbf{G} \cdot \begin{bmatrix} 100 \\ 010 \\ 010 \\ 001 \end{bmatrix}}_{\mathbf{G'}} \cdot \underbrace{\begin{bmatrix} j_{11}(z) \\ j_{21}(z) \\ j_{22}(z) \end{bmatrix}}_{\overrightarrow{\mathbf{J'}_S}(z)} = \mathbf{G'} \cdot \overrightarrow{\mathbf{J'}_S}(z). \tag{4}$$

When the column rank of the modified sensing matrix is three—i.e., when $\mathbf{G'}^{\dagger} \cdot \mathbf{G'}$ is invertible—the vectorized Jones matrix of the sample can be calculated using the pseudoinverse as

$$\overrightarrow{\mathbf{J'}_S}(z) = \left(\mathbf{G'}^{\dagger} \cdot \mathbf{G'}\right)^{-1} \cdot \mathbf{G'}^{\dagger} \cdot \mathbf{m}(z). \tag{5}$$

To achieve this condition, at least three tomographic measurements acquired with different waveplate orientations are required to enable unambiguous reconstruction of the sample's Jones matrix.

### D. Phase Correction

Pathlength drift between the sample and reference arms introduces phase variations between repeated measurements of the complex-valued tomograms [25]. Our strategy for estimating these phase shifts is to perform phase optimization by minimizing the disagreement among repeated measurements obtained from more than three waveplate orientations. Note that typical phase variations are depth-independent; however, regions near vessels exhibit depth-dependent phase variations, requiring depth-dependent phase compensation. Details are provided in the Discussion section.

Each acquired A-line measurement deviates from the ideal signal by an unknown phase shift combined with random noise. This effect can be modeled as

$$\tilde{\mathbf{m}} = \mathrm{diag}\left(e^{-i\vec{\boldsymbol{\varphi}}}\right) \cdot \mathbf{m} + \boldsymbol{\eta} = \boldsymbol{\Phi}^{\dagger} \cdot \mathbf{m} + \boldsymbol{\eta}, \quad (7)$$

where $\vec{\boldsymbol{\varphi}}$ is a vector of unknown phase offsets (one per A-line measurement), $\boldsymbol{\Phi}$ is the corresponding diagonal phase matrix, and $\boldsymbol{\eta}$ is a noise matrix that perturbs the ideal measurements. Without loss of generality, the first element of $\vec{\boldsymbol{\varphi}}$ is set to zero to define the global phase reference.

First, we compensate for the apparent phase differences $\tilde{\boldsymbol{\Phi}}$ at the sample surface across repeated A-line measurements

$$\tilde{\mathbf{m}}_0 = \tilde{\boldsymbol{\Phi}} \cdot \tilde{\mathbf{m}}. \quad (8)$$

We refer to this procedure as zero-phase correction, which eliminates random phase noise but also the meaningful phase offsets arising from tissue retardance along the light path, such as the cornea in retinal imaging.

Second, we recover the meaningful phase offsets through optimization. From Eqs. (5) and (8), the sample's Jones matrix can be recovered as

$$\widehat{\widetilde{\mathbf{J}'_S}} = \left(\mathbf{G}'^{\dagger} \cdot \mathbf{G}'\right)^{-1} \cdot \mathbf{G}'^{\dagger} \cdot \boldsymbol{\Phi}_0 \cdot \tilde{\mathbf{m}}_0, \quad (9)$$

where $\boldsymbol{\Phi}_0$ represents an estimate of the phase offsets arising from tissue birefringence along the light path. Any phase estimation vector results in a plausible Jones matrix. The most suitable phase compensation is obtained by minimizing the reconstruction error of the corresponding Jones matrix

$$\hat{\boldsymbol{\Phi}} = \arg\min_{\boldsymbol{\Phi}_0} \left\| \boldsymbol{\Phi}_0 \cdot \tilde{\mathbf{m}}_0 - \mathbf{G}' \cdot \widehat{\widetilde{\mathbf{J}'_S}} \right\|_2^2 \quad (10)$$

This phase estimation can be transformed into

$$\begin{aligned} \hat{\boldsymbol{\Phi}} &= \arg\min_{\boldsymbol{\Phi}_0} \|\mathbf{U} \cdot \boldsymbol{\Phi}_0 \cdot \tilde{\mathbf{m}}_0\|_2^2 \\ &= \arg\min_{\boldsymbol{\Phi}_0} \vec{\boldsymbol{\Phi}}_0^{\dagger} \cdot \mathbf{W} \cdot \vec{\boldsymbol{\Phi}}_0, \end{aligned} \quad (11)$$

where $\mathbf{U} = \boldsymbol{I} - \mathbf{G}' \cdot \left(\mathbf{G}'^{\dagger} \cdot \mathbf{G}'\right)^{-1} \cdot \mathbf{G}'^{\dagger}$, $\mathbf{W} = \mathrm{diag}(\tilde{\mathbf{m}}_0)^{\dagger} \cdot \mathbf{U}^{\dagger} \cdot \mathbf{U} \cdot \mathrm{diag}(\tilde{\mathbf{m}}_0)$, and $\vec{\boldsymbol{\Phi}}_0$ designates the vectorized diagonal elements of $\boldsymbol{\Phi}_0$. The matrix $\mathbf{W}$ is determined jointly by the tomographic measurements and the waveplate orientations, owing to $\boldsymbol{\Phi}_0 \cdot \tilde{\mathbf{m}}_0 = \mathrm{diag}(\tilde{\mathbf{m}}_0) \cdot \vec{\boldsymbol{\Phi}}_0$. Because $\mathbf{W}$ is non-coherent, it can be averaged over a suitable depth range and across multiple A-lines. This optimization can be performed using gradient descent, where the phase gradient at iteration $n$ is computed as

$$\Delta\vec{\boldsymbol{\Phi}}_n = \Re\left\{ i\, \boldsymbol{\Phi}_n{}^{\dagger} \cdot \mathbf{W} \cdot \vec{\boldsymbol{\Phi}}_n \right\}. \quad (12)$$

Empirically, the optimization exhibits four local minima, two of which correspond to unrealistically high retardance and diattenuation and can therefore be excluded. The remaining two solutions exhibit similar retardance values but optic axis orientations separated by $\pi$. To resolve this ambiguity, we enforce optic axis consistency across neighboring A-lines and B-scans. Of the two possible frame-consistent solutions, we select the one with the smaller, we select one with the smaller residual errors at the minima. The chosen axis orientation then determines the corresponding phase offsets used for correction.

### *E. Coherency Filtering*

Coherency filtering aims to retain the spatial coherence properties of the OCT signal while suppressing noise and coherent artifacts. A stack of RevoPol measurements acquired at different waveplate orientations at each pixel can be described as a complex vector $\mathbf{m}$. The local coherency matrix is computed as the outer product

$$\mathbf{H} = \mathbf{m}\mathbf{m}^{\dagger}, \quad (13)$$

which is a rank-1 Hermitian matrix. To suppress uncorrelated noise, the coherency matrices are spatially filtered using a Gaussian kernel (11 × 11, axial × lateral, σ = 4) to obtain the matrix $\mathbf{H}_{\mathrm{filt}}$, which is no longer rank-1 yet still Hermitian. The filtered coherency matrix is then decomposed by singular value decomposition,

$$\mathbf{H}_{\mathrm{filt}} = \mathbf{V}\mathbf{S}\mathbf{V}^{\dagger}. \quad (14)$$

Only the dominant coherent mode is retained, and a denoised measurement vector is reconstructed as

$$\mathbf{m}_{\mathrm{filt}} = \sqrt{\mathbf{s}_1}\mathbf{v}_1. \quad (15)$$

where $\mathbf{v}_1$ is the leading singular vector and $\mathbf{s}_1$ is its singular value, which encodes the coherent signal power. This corresponds to the least squares approximation of $\mathbf{H}_{\mathrm{filt}}$ with a rank-1 matrix.

We define a metric as the normalized ratio of the dominant singular value to the total energy,

$$Q = \frac{\mathbf{s}_1}{\sum_i \mathbf{s}_i}, \quad (16)$$

which delineates tissue signal with a high $Q$ from noise-dominated or dynamically changing areas featuring a low $Q$. A threshold ($Q > 0.55$) is applied to extract retinal signals starting from the internal limiting membrane. Small isolated regions are removed using morphological filtering, and the detected surface profiles are further smoothed with a 25-pixel median filter to ensure continuity. In addition, the $Q$ metric enables filtering of shadowing artifacts from large retinal vessels.

### *F. Imaging Workflow*

This study involved simulations, imaging of birefringent phantoms, and retinal imaging in healthy human eyes. Simulations were conducted to validate the RevoPol framework, while phantom experiments were used to calibrate the waveplate's rotation speed and confirm robustness *ex vivo*. The phantoms were fabricated from polycarbonate embedded in an epoxy matrix, as previously described [26], and an additional, static waveplate was inserted into the imaging path to simulate corneal birefringence. To demonstrate applicability *in vivo*, retinal imaging was performed in a healthy volunteer. The pilot human study adhered to the tenets of the Declaration of Helsinki and was approved by the Mass General Brigham Institutional Review Board (IRB). Written informed consent was obtained from all participants prior to enrollment. The following section details the procedures used to extract tissue polarization properties with the RevoPol framework.

Step 1: Image acquisition. The waveplate rotation speed was set to approximately $\pi/12$ radians between each repeated B-scan. For a B-scan pattern comprising 1024 A-lines (20° field-of-view) at an A-line rate of 85 kHz, this corresponded to a rotation speed of approximately 3 revolutions per second. The actual rotation speed was calibrated by imaging a birefringent phantom and analyzing the periodicity of the resulting signal together with the uniformity of the reconstructed optic axis orientation. While the waveplate was rotating, six tomograms were acquired at the same location using the system's OCTA mode to enable reconstruction of the Jones matrices for a cross-sectional image.

Step 2: Image pre-processing. Image registration was performed to align repeated A-line measurements, reducing motion artifacts and ensuring signal consistency for accurate phase analysis. For axial registration, correlations were evaluated between repeated spectra after lateral Hann-window filtering, whereas lateral registration was performed using local windows with 50% overlap. The correlation profiles were evaluated using the chirp z-transform, enabling efficient subpixel estimation of axial and lateral displacements within predefined search ranges. The peak correlation positions and corresponding correlation magnitudes were identified for all scan pairs. Global axial and lateral shift profiles were then estimated by fitting third-order polynomial models to all pairwise peak positions, weighted by the corresponding correlation magnitudes. Coherency filtering and layer-specific segmentation (e.g., retinal surface or retinal pigment epithelium) were subsequently applied to the registered tomograms.

Step 3: Phase correction. For the target layer, zero-phase correction was applied based on layer-specific segmentation, eliminating both random phase noise and systematic phase offsets induced by birefringent media (e.g., the cornea in retinal imaging) along the light path. To estimate the phase offset, the W matrix from (11) was constructed by averaging signals along depth over several pixels. After excluding unrealistically high retardance and diattenuation solutions, the phase optimization approach would yield two possible phase offsets. Neighboring A-lines and B-scans were grouped according to local continuity of the recovered optic axis orientations. The group with the smaller overall residual error was selected and used to determine the phase offset. The phase offset estimation step can be performed in a depth-dependent fashion or skipped if no birefringent media is present along the imaging path beyond the target surface.

Step 4: Polarimetric reconstruction. The estimated phase offsets were applied to the filtered tomograms, followed by reconstruction of the complex-valued Jones matrix representing the polarization response of the sample. The Jones matrix was then decomposed into a unitary rotation matrix, from which the cumulative retardance and optic axis orientation were extracted. The optic axis orientations were compensated for the waveplate rotation, determined by the rotation speed, the inter-B-scan time interval, and the B-scan timestamps. Once the cumulative Jones matrix was determined, local depth-resolved retardance and optic axis orientation could be derived, although this was only done for phantom imaging [23].

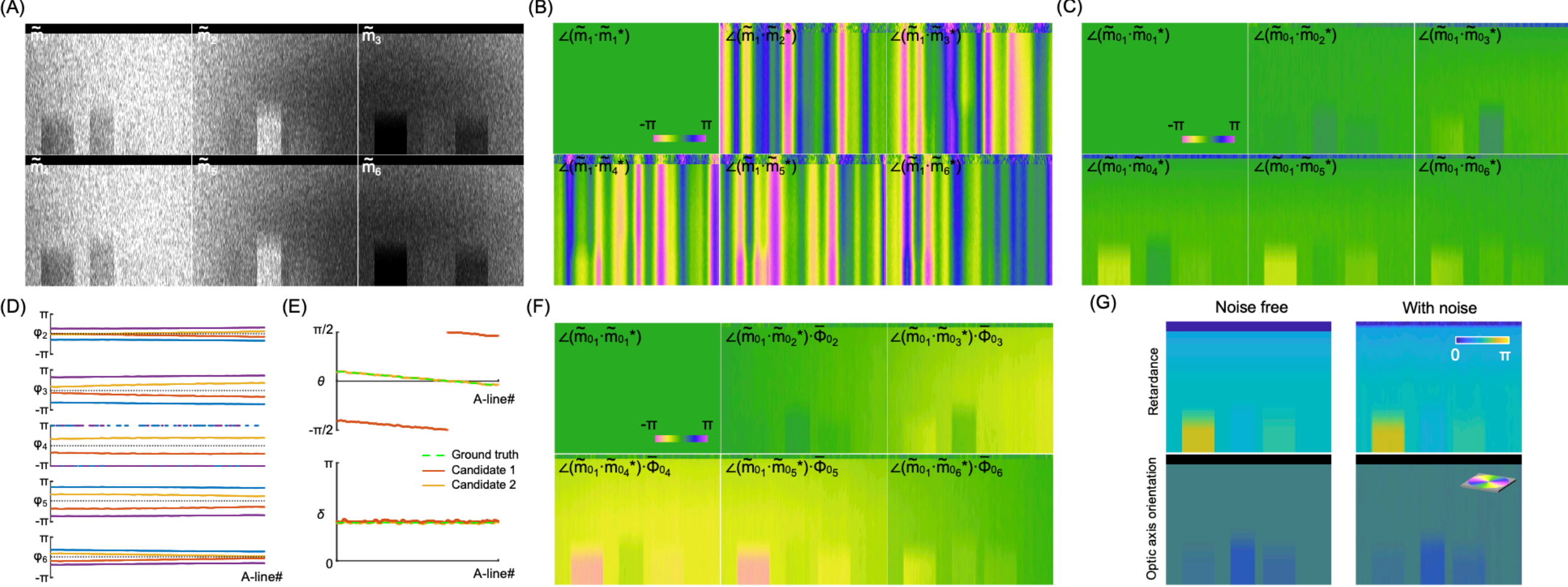

Fig 3. Simulations of imaging and polarimetric reconstruction using the RevoPol framework. (A) Intensity of sequential tomographic measurements acquired during waveplate rotation (rotation step: π/12 per B-scan) with speckle and phase noise applied. Three birefringent bars and surface cumulative retardance were included in the simulation. (B) Phase differences between the first and subsequent measurements. (C) Phase differences after zero-phase correction. (D) Four candidate phase-offset solutions ($\varphi_1 = 0$) obtained from phase optimization. The blue and purple curves correspond to solutions with unrealistically high retardance and diattenuation. The yellow curves indicate the selected solution with the smallest residual error, whereas the orange curves represent the rejected candidate solution. (E) Two candidate solutions (colors corresponding to D) recovering the optic axis orientation (θ) and retardance (δ), compared with the ground-truth retarder (green dashed line). (F) Final corrected phase differences obtained by applying the phase offsets from (D) to (C). (G) Cumulative retardance and optic axis orientation reconstructed from tomograms with corrected phases, compared with polarimetric reconstruction from noise-free signals.

Step 5: Surface retardance compensation. Surface retardance, originating from birefringent tissues along the imaging path (e.g., the cornea [27]), was extracted as rotation vectors. The square root of these vectors was converted into Jones matrices and applied symmetrically to the measured Jones matrices, thereby removing the cumulative retardance contribution along the imaging path [21].

### G. Depth-multiplexed PS-OCT system

A depth-multiplexed PS-OCT system was used as a reference for comparison with RevoPol and has been described previously [28]. Briefly, the system employed two orthogonal polarization-diverse detection channels, enabling simultaneous acquisition of the full Jones matrix at each spectral point of every A-line. The system operated with an Axsun wavelength-swept laser source (Excelitas Technologies Corp., USA) centered at 1060 nm with a 100 kHz A-scan rate and was integrated with a Spectralis HRA+OCT platform (Heidelberg Engineering, Germany).

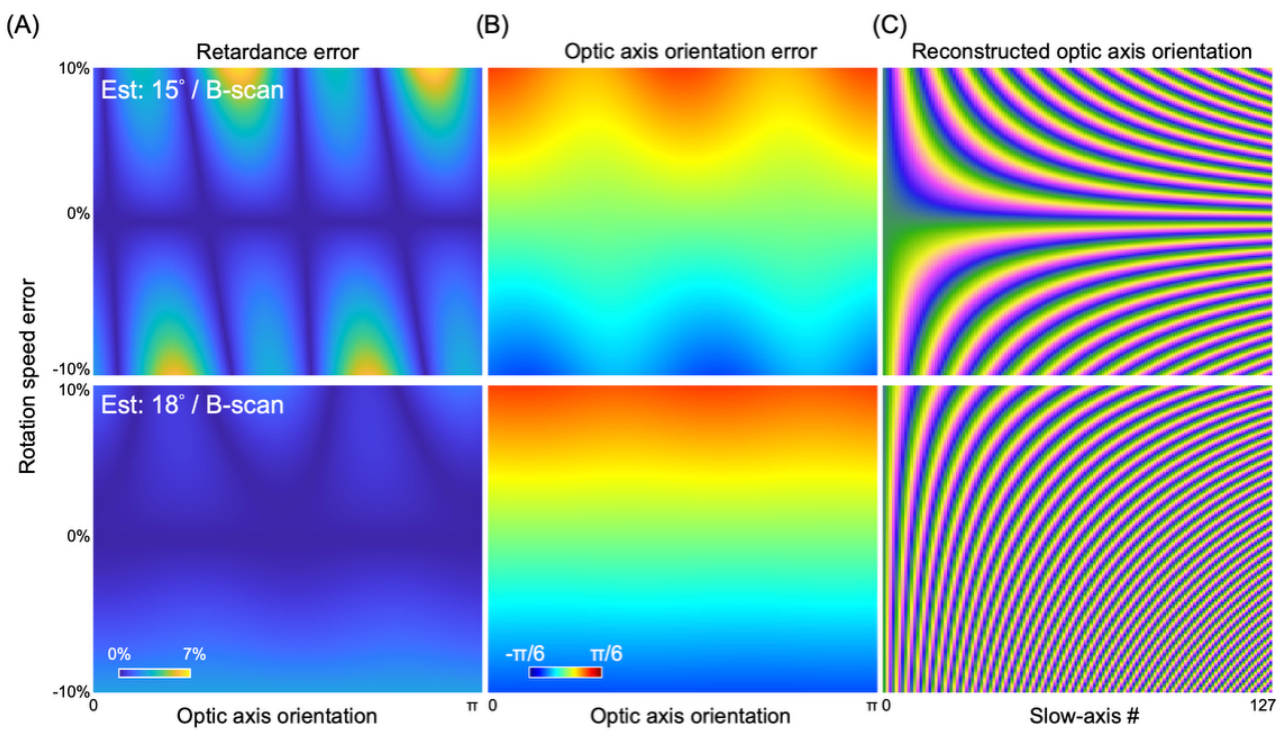

Fig. 4. Simulation of rotation speed error on polarimetric reconstruction. Estimated rotation speeds (ω) are 15°/B-scan (top) and 18°/B-scan (bottom). Rotation speed error is defined as $(\omega_{actual} - \omega_{estimated})/\omega_{estimated} \times 100\%$. (A) Retardance error and (B) optic axis orientation error as a function of known optic axis orientation. (C) Reconstruction of a constant optic axis orientation along the B-scan direction without compensation for continuous rotation. With an estimated speed of 15°/B-scan, the orientation remains constant when the rotation speed is correct (Rotation speed error = 0%) and exhibits progressive variation in the presence of speed error. In contrast, an estimated speed of 18°/B-scan produces high-frequency fluctuations in

## III. Results

### A. Simulation

We used numerical simulations to assess the ability of the RevoPol framework to reconstruct polarization properties. The simulated dataset included speckle and phase noise, as well as known birefringent structures consisting of three embedded bars and significant cumulative surface retardance. Sequential tomographic measurements were generated during waveplate rotation, and phase differences were computed relative to the initial acquisition (**Fig. 3A–B**). With the waveplate increment of π/12 radians between repeated B-scans, the first–fourth, second–fifth, and third–sixth measurements exhibited matching intensities but showed distinct phase differences. Zero-phase correction effectively removed random phase noise (**Fig. 3C**). Phase optimization then produced four candidate phase-offset solutions (**Fig. 3D**). After excluding two solutions associated with unrealistically high retardance and diattenuation, the remaining candidate with the smaller residual error matched the ground-truth retarder (**Fig. 3E**). Applying the recovered phase offsets enables accurate reconstruction of cumulative retardance and optic axis orientation (**Fig. 3F–G**). We evaluated the effect of rotation speed estimation on polarimetric reconstruction. We found that an estimated speed of 18° (π/10) per B-scan was the most robust against errors in reconstructed retardance and optic axis orientation (**Fig. 4A–B**). In comparison, at 15° (π/12) per B-scan the reconstructed orientation remained largely stable along the slow-axis scan, showing only minor drifts in a C-scan under estimation error when continuous rotation was not compensated (**Fig. 4C**).

### B. Phantom Imaging

We applied the RevoPol framework to a birefringent phantom with a retarder placed in the imaging path to experimentally validate the reconstruction workflow, using a waveplate rotation speed of π/12 (~15°) per B-scan (**Fig. 5A-F**). Phase shifts were uniform along each A-line but varied randomly between successive A-lines (**Fig. 5B**). Using the phase optimization approach, we obtained optic axis orientation and retardance at the surface, where the apparent optic axis orientation varied with waveplate rotation while the retardance remained constant (**Fig. 5C**). Using the correct solution, consistent phase offsets were recovered, which enabled reconstruction of cumulative retardance and optic axis orientation across the phantom (**Fig. 5D–E**). Polarization properties were further corrected through surface retardance compensation, ensuring that the reconstructed properties originated from the phantom itself (**Fig. 5F**). We measured cumulative retardance at the surface using different waveplate rotation speeds (**Fig. 5G**). Both measurements exhibited consistent patterns, demonstrating that the polarimetric reconstruction was independent of the waveplate speed.

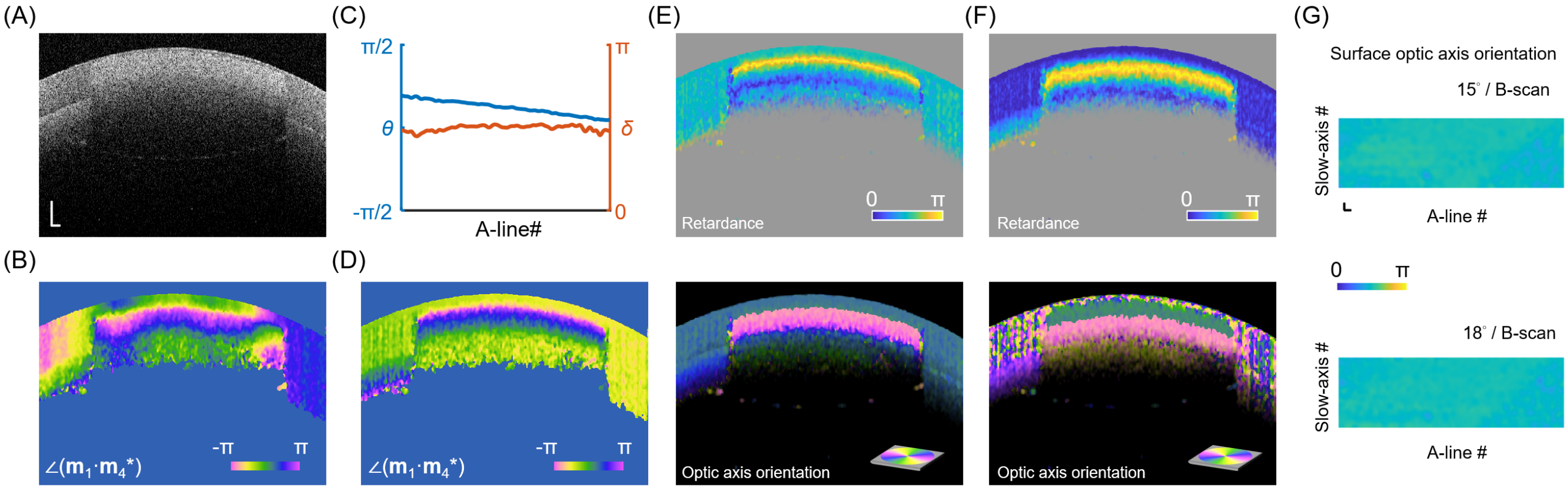

Fig 5. Imaging of a birefringent phantom using the RevoPol framework. (A) Structural B-scan. (B) Phase difference map before correction. (C) Optic axis orientation (*θ*) and retardance (*δ*) estimated via phase optimization. (D) Phase difference map after correction. (E) Cumulative retardance and optic axis orientation reconstructed from corrected tomograms. (F) Cumulative retardance and optic axis orientation after surface retardance compensation. (G) Surface retardance maps acquired using different waveplate rotation speeds. All scale bars: 200 μm.

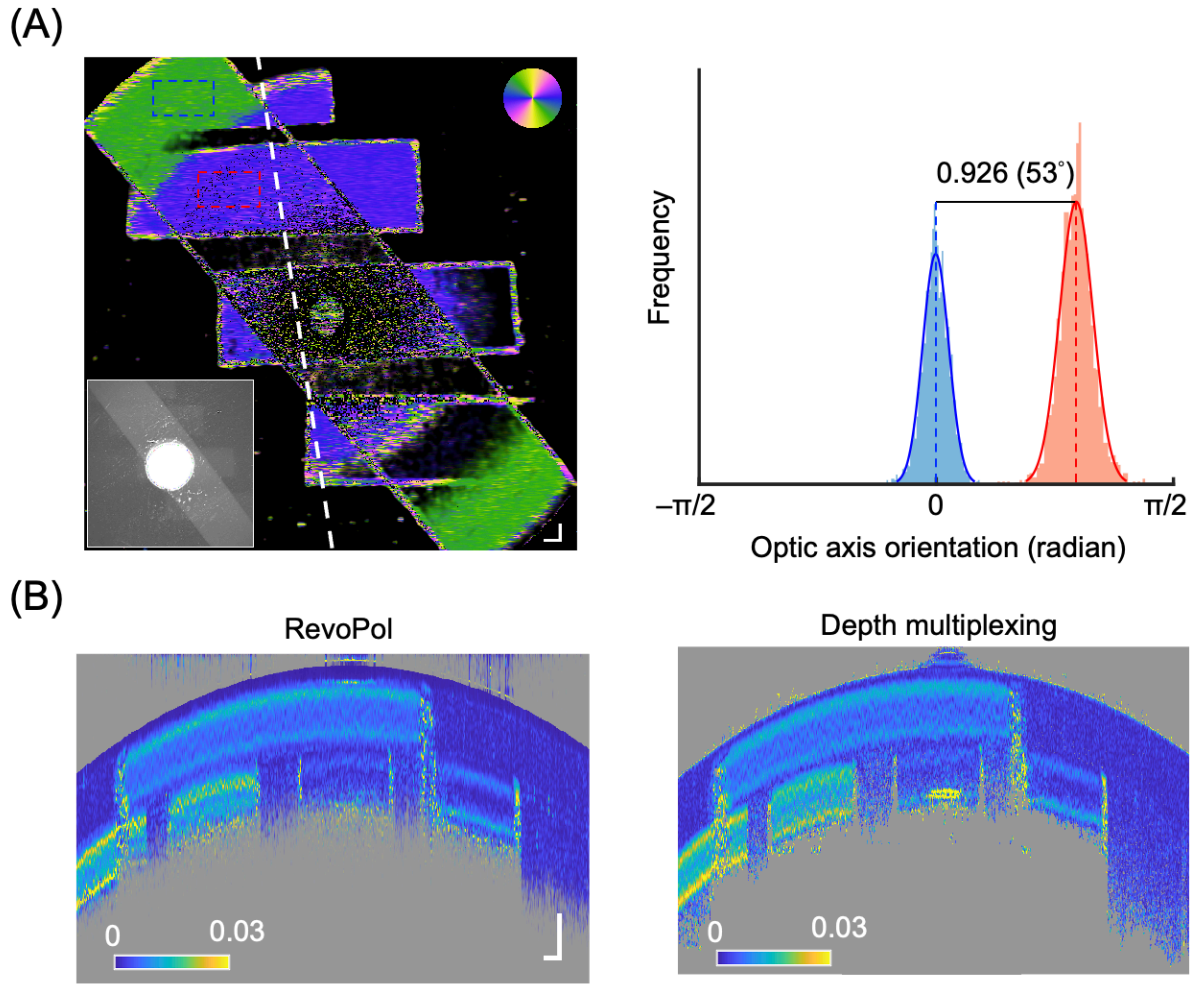

Fig. 6. Validation polarimetric reconstruction using the RevoPol framework. (A) Left: En face map of local optic axis orientation in phantom strips, alongside the scanning laser ophthalmoscopy (SLO) preview. The central region exhibits strong reflection artifacts. Right: Distributions of optic axis orientations within the blue and red dashed regions. (B) Comparison of local retardance measured by RevoPol and a conventional depth-multiplexing PS-OCT system, along the white dashed line in (A).

We quantified the depth-resolved optic axis orientations of the phantom strips and observed a difference of 53° between the two layers (**Fig. 6A**), which is consistent with the known physical alignment. In addition, we compared the depth-resolved retardance profiles between RevoPol and depth-multiplexed PS-OCT (**Fig. 6B**). After accounting for differences in wavelength and axial sampling, the two methods show strong agreement with local retardance measurements, supporting the accuracy of the RevoPol reconstruction framework.

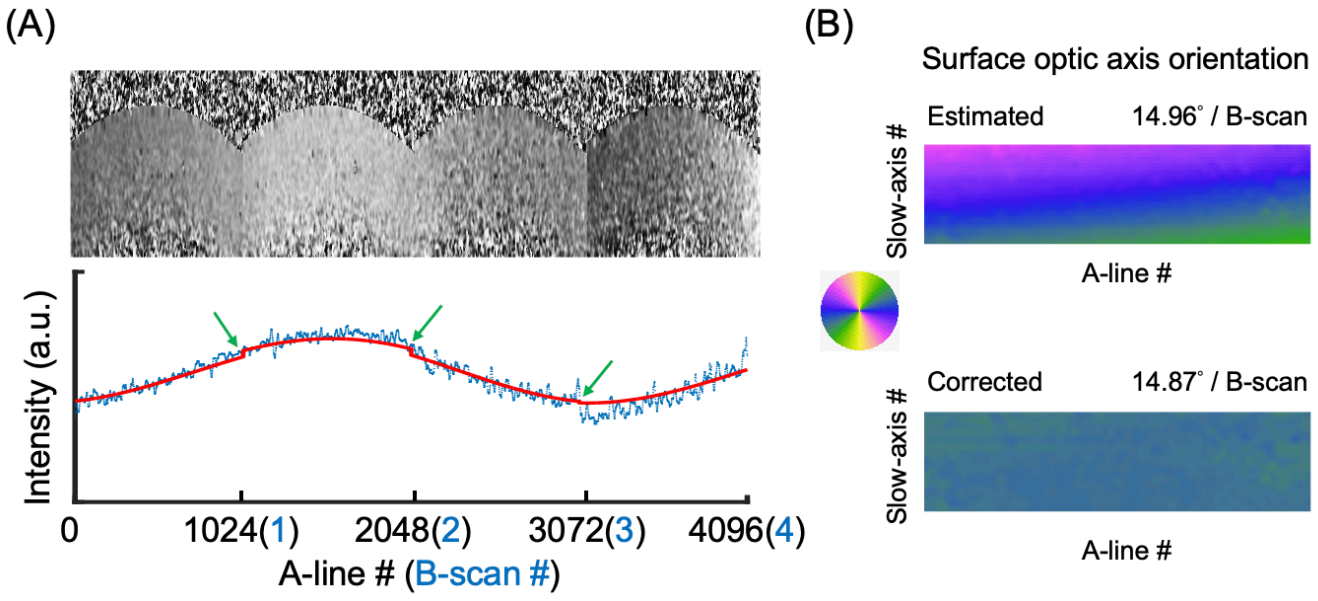

Fig. 7. Calibration of waveplate rotation speed. (A) During waveplate rotation, the fixed retarder placed along the light path induced periodic intensity modulation in the phantom. After coherence filtering, the normalized intensity across repeated B-scans (blue) shows a periodicity of 3 B-scans, corresponding to a waveplate rotation speed of π/12 (~15°) per B-scan. The intensity inconsistency between B-scans (green arrows) indicates inter-B-scan time. (B) In a C-scan, the reconstructed surface optic axis orientation shows drifts (top) due to small deviations in the estimated rotation speed. Precise calibration of the waveplate rotation speed eliminates these drifts (bottom).

We used a non-birefringent phantom with a retarder placed in the imaging path (between the lens and the phantom) to calibrate the waveplate rotation increment per B-scan (**Fig. 7**). For a rough estimation, we analyzed the signal intensity modulated by a fixed retarder along the imaging path (**Fig. 7A**). After applying coherence filtering, the normalized intensity across repeated B-scans exhibited a periodicity of three B-scans. The intensity varies as

$$I = |m|^2 \propto \cos(8\psi), \tag{17}$$

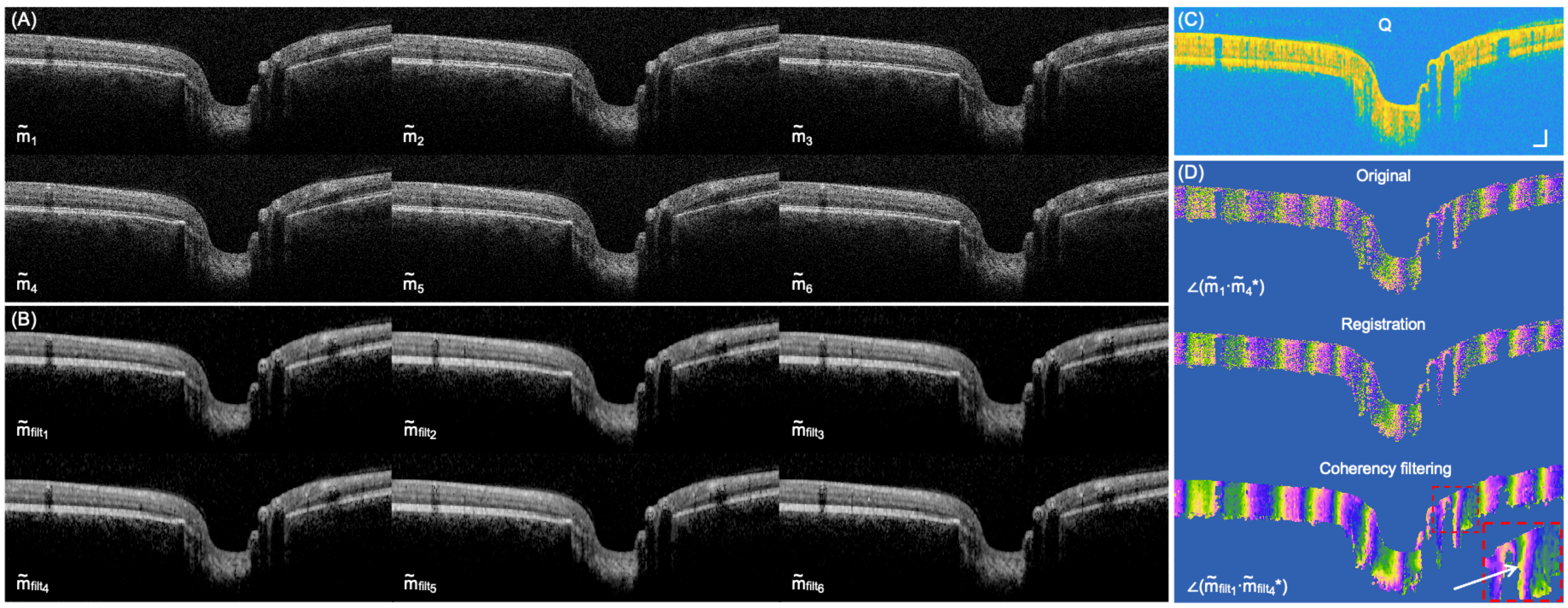

Fig. 8. Effect of coherency filtering on tomograms. (A) Original tomograms. (B) Tomograms after registration and coherency filtering, shown without phase correction. (C) $Q$ metric derived from the coherency matrix, used to enhance boundary contrast. (D) Phase difference between the first tomogram and a repeated measurement for the original, registered, and coherency-filtered data, masked by $Q > 0.55$. Registration reduces phase noise, and coherency filtering further suppresses speckle and phase fluctuations while preserving structural boundaries. The white arrow indicates abrupt phase changes within single A-lines in regions near blood vessels. Scale bar: 200 μm.

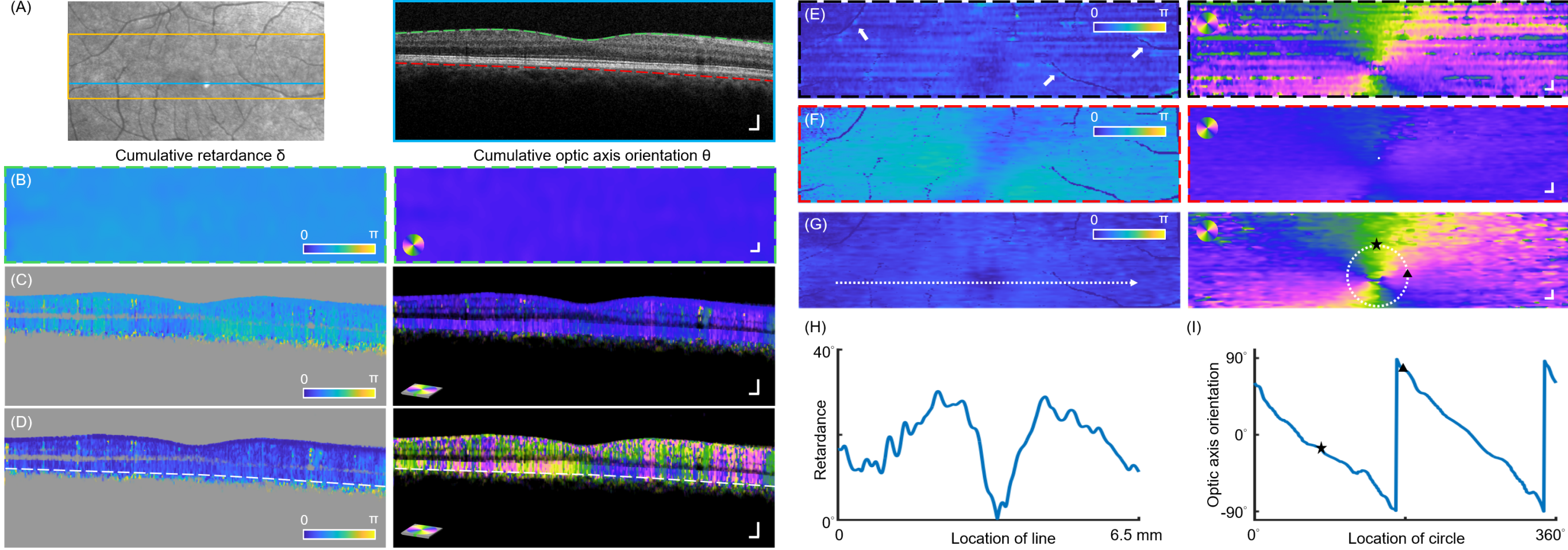

Fig. 9. Polarimetric reconstruction of the human retina using RevoPol. (A) Left: Scanning laser ophthalmoscopy (SLO) image with the scan region (orange box) centered on the fovea. Right: Corresponding B-scan intensity along the blue line in the SLO image. (B) En face maps of surface retardance (δ, 60.1° ± 3.0°) and optic axis orientation (θ, 100.7° ± 2.7°), within the region indicated by the green line in (A). (C) B-scan cumulative retardance and optic axis orientation before surface retardance compensation. (D) B-scan cumulative retardance and optic axis orientation after surface retardance compensation. (E) En face cumulative retardance and optic axis orientation extracted along the white lines in (D). White arrows indicate retardance artifacts. (F) En face retardance map estimated via phase correction at the depth corresponding to the red line in (A). (G) En face retardance map after surface retardance compensation, incorporating the estimate in (F). (H) Retardance profile along the white dotted line in (G). (I) Optic axis orientation profile along the white circular path in (G). All scale bars: 200 μm.

where $\psi$ is the evolving half-waveplate fast-axis angle, i.e., the angular period of the waveplate is one-eighth of the observed intensity period. This yielded a waveplate rotation speed of $\pi/12$ (~15°) per B-scan. For accurate calibration, the reconstructed surface optic axis orientations in a C-scan are highly sensitive to rotation speed (**Fig. 7B**), making uniform orientation across the scan a reliable metric for precise waveplate calibration.

## C. Retinal Imaging

In retinal imaging, the original tomograms exhibited pronounced speckle patterns on the retinal structures (**Fig. 8A**). After registration and coherency filtering, speckle noise was substantially suppressed while structural boundaries were preserved (**Fig. 8B**). The $Q$ metric further highlights retinal boundaries and effectively excludes regions shadowed by blood flow (**Fig. 8C**). The corresponding phase-difference maps (**Fig. 8D**) show that the original data contain significant phase noise within A-lines. Registration reduces this noise, and coherency filtering further suppresses high-frequency fluctuations while retaining phase-difference features, i.e., the phase variations across A-lines. It is worth noting that these A-line phase fluctuations are more pronounced in retinal data than in phantom experiments, primarily due to eye motion *in vivo* compared with the stationary phantom. We further demonstrated the feasibility of RevoPol for polarimetric reconstruction in the healthy human retina (**Fig. 9**). In the fovea-centered scan (**Fig. 9A**), the surface retardance and optic axis orientation appeared spatially uniform (**Fig. 9B**), suggesting that corneal birefringence along the light path can be effectively modeled as a well-aligned retarder. Before surface retardance compensation, the cumulative polarimetric results are dominated by surface contributions (**Fig. 9C**). After compensation, the underlying retinal polarimetric features become visible (**Fig. 9D**). However, directly extracting the en face retardance map from the C-scan after surface compensation reveals artifacts near blood vessels (**Fig. 9E**), likely associated with abrupt phase changes observed in these regions (**Fig. 8D**). Instead, we performed phase correction at the retinal pigment epithelium layer (**Fig. 9F**), coinciding with the depth for extracting surface cumulative retardance, which resulted in more robust reconstruction of cumulative polarimetric properties (**Fig. 9G**). We observed reduced retardance near the foveal center, with maximum and minimum values of approximately 30° and 0.62°, respectively (**Fig. 9H**). In addition, a characteristic radial optic axis orientation was observed, corresponding to photoreceptor axons extending outward from the fovea (**Fig. 9I**). These birefringence features are in close agreement with established descriptions of Henle's fiber layer [21, 29, 30]. Then, we demonstrated RevoPol polarimetric and angiographic reconstruction in the healthy human optic nerve head (**Fig. 10**). After compensating for surface retardance, the cumulative retardance exhibited a circumpapillary annular pattern consistent with the retinal nerve fiber layer thickness distribution (**Fig. 10A–B**). Importantly, OCTA reconstructions using RevoPol and the standard system showed no significant differences (**Fig. 10C**). The clear visualization of retinal birefringence and organization highlights the ability of RevoPol to recover physiologically relevant polarization signatures *in vivo* with minimal hardware modification to the clinical OCT platform.

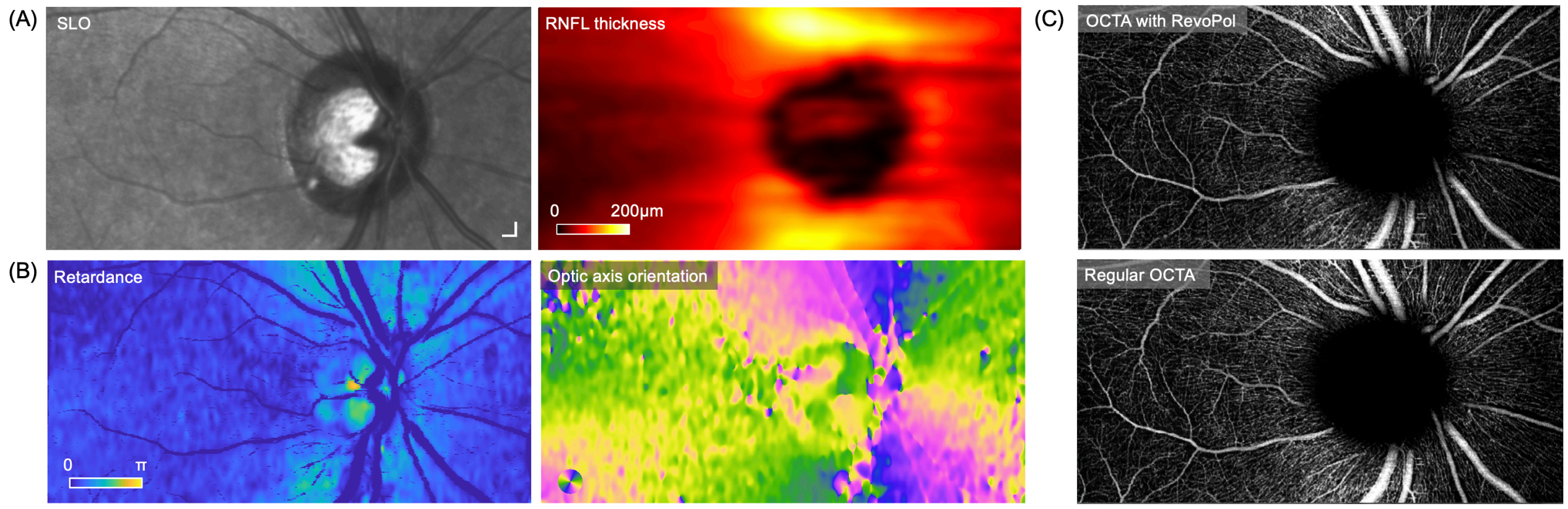

Fig. 10. Polarimetric and angiographic reconstruction of healthy human optic nerve head using RevoPol (A) Left: Scanning laser ophthalmoscopy (SLO) image. Right: Retinal nerve fiber layer (RNFL) thickness measured using Heidelberg Eye Explorer. (B) The cumulative retardance and optic axis orientation maps at the retinal pigment epithelium, with vessel regions excluded. (C) Comparison of OCTA reconstructions from the inner limiting membrane (ILM) to the inner plexiform layer (IPL) obtained using RevoPol and the standard system configuration, processed and visualized with Heidelberg Eye Explorer. All scale bars: 200 μm.

## IV. Discussion

In this study, we presented RevoPol, a new PS-OCT framework that enables full Jones matrix reconstruction using a single-detector OCT system. As a practical implementation, we developed a detachable rotating waveplate module that can be integrated into a commercial OCT platform. Using this setup, we demonstrated reconstruction of retardance and optic axis orientation in both birefringent phantoms and the healthy human retina. These results confirm that RevoPol achieves effective reconstruction of known polarization signatures *in vivo* through a simple add-on module compatible with existing OCT hardware. The concept underlying RevoPol shares similarities with earlier approaches to scanning laser polarimetry [31]. The modular strategy of introducing controlled polarization-state modulation through an external add-on component may also inspire implementations in other polarization-sensitive imaging modalities, such as polarized light microscopy (PLM) [32], polarization-resolved second harmonic generation (P-SHG) microscopy [33], and dichroism-sensitive photoacoustic imaging [34], although the underlying contrast mechanisms and reconstruction methods differ substantially among these techniques.

Our study demonstrates the implementation of the polarization-sensitive module on the SPECTRALIS platform. Compatibility with other commercial OCT systems depends on the mechanical design, optical configuration, and accessibility of acquisition data from the host instrument. In general, integration of RevoPol with commercial OCT platforms requires the following conditions: (a) Unidirectional repeated B-scan capability: Bidirectional scanning would introduce non-uniform waveplate angular increments between consecutive B-scans. (b) Access to raw OCT signal: RevoPol reconstruction requires complex-valued OCT data (both amplitude and phase). (c) Access to B-scan timestamps: Although hardware synchronization is not required, accurate B-scan timestamps are essential for estimating the optic axis orientation of the rotating waveplate for each acquisition. (d) Controlled input polarization state: The input polarization state should be linear. If polarization controllers are integrated into the system, they need to be adjusted accordingly. (e) Stable and repeatable scanning: For imaging *in vivo*, motion stability (e.g., eye tracking) is important to ensure consistency across repeated B-scans. This requirement is relaxed for static samples.

Compared with existing PS-OCT methods, RevoPol offers distinct advantages. A previous study employed a linear polarizer followed by an LCD-based addressable wave plate set to 45°, 90°, and 135° [35]. This configuration required longer imaging times and lacked phase correction, which prevented its use for *in vivo* imaging. Conventional PS-OCT with a single circular input state is limited to reporting cumulative retardance or assuming a uniform optic axis orientation with depth [36, 37]. This limitation arises from an intrinsic degeneracy in the polarization measurement: whenever the detected polarization state coincides with the mirror state of the input—that is, the same state with flipped helicity reflected across the horizontal QU-plane on the Poincaré sphere—the reconstruction becomes indeterminate [38, 39]. For example, with a perfectly circular input state and an observed circular state of opposite handedness, the measured retardance is π, yet the optic axis orientation cannot be determined. In contrast, RevoPol enables full Jones matrix reconstruction, allowing calculation of both depth-resolved birefringence and optic axis orientation. While multi-input PS-OCT offers robust polarimetric reconstruction, its implementation typically relies on depth-multiplexing [40, 41] or electro-optic polarization modulators [42] in combination with polarization-diverse detection. These requirements substantially increase system complexity and have limited widespread translational and clinical adoption. In contrast, RevoPol employs sequential polarization modulation with a rotating waveplate, combined with a sensing matrix and phase optimization, to recover tissue polarization properties using standard OCT hardware. This framework preserves the integrity of the underlying OCT system. Its modular design supports clinical translation by lowering the barriers to implementation and maintenance, particularly in multi-site studies where hardware standardization is critical. We acknowledge that RevoPol requires repeated B-scans for reconstruction, thereby reducing effective imaging speed and increasing sensitivity to motion. Although these represent key limitations relative to conventional multi-input PS-OCT, RevoPol naturally integrates with OCTA acquisition schemes, thereby minimizing its impact on clinical workflow.

Despite the demonstration here using spectral-domain OCT, no fundamental limitation is expected for swept-source implementations. The optic axis variation of the rotating waveplate within a single A-line acquisition is negligible, as the A-line dwell time (μs scale) is orders of magnitude shorter than the waveplate rotation period (~100 ms). Therefore, the polarization state can be considered effectively constant during each wavelength sweep. Swept-source OCT systems, which typically operate at longer center wavelengths, may further provide improved penetration depth. In retinal imaging, combining RevoPol with swept-source OCT may enable characterization of deeper structures such as the sclera and lamina cribrosa.

The choice of waveplate rotation speed plays an important role in the accuracy and practicality of RevoPol reconstruction. Ideally, any rotation speed would be sufficient for polarimetric reconstruction within the RevoPol framework. In practice, however, the selected speed affects reconstruction stability and quality control. From simulations with six repeated measurements, we found that a $\pi/10$ rotation step provides the greatest robustness against reconstruction errors when the estimated rotation speed deviates from the actual speed. It is worth noting that this configuration does not yield the lowest condition number of the sensing matrix G, and the reason for its robustness remains unclear. For retinal imaging, we nevertheless implemented six repeated measurements with a $\pi/12$ rotation step. Under this configuration, the initial optic-axis orientations of the input light remain approximately constant for the first B-scan along the slow-axis direction. This property facilitates fine-tuning of the rotation-speed estimation, ensuring that the reconstructed optic-axis orientations of the surface remain approximately constant along the slow-axis direction, particularly when imaging timestamps are inaccurate or unavailable. Accurate estimation of the rotation speed further reduces reconstruction errors. Although a $\pi/12$ rotation step for reconstruction is not optimal if the estimated speed deviates from the true speed, the measured rotation speed variation in our module is below 1%, indicating that the resulting motion jitter introduces only a negligible error. In addition, the first–fourth, second–fifth, and third–sixth measurements correspond to orthogonal polarization states and therefore exhibit matching intensities. This feature may benefit angiographic reconstruction and provide a convenient quality-control metric for active eye tracking.

The current, unsynchronized setup simplifies implementation but introduces two limitations: (i) because only an averaged rotation speed is estimated, intra-volume speed fluctuations may be missed, potentially introducing residual phase errors; and (ii) the initial optic axis orientation of the rotating waveplate is unknown, complicating cross-dataset comparisons, though retardance estimates remain unaffected. In retinal imaging, the characteristic optic axis orientation of the Henle fiber layer may serve as a reference for comparing absolute optic-axis orientations across datasets. A straightforward improvement would be to record the real-time orientation of the waveplate for each B-scan and align it with OCT acquisition timing. In contrast, directly driving the rotation stage using OCT timing signals may lead to non-uniform angular velocity due to motor and waveplate inertia, thereby violating the assumption of constant rotation.

Phase correction is critical for accurate polarimetric reconstruction in the RevoPol framework. Phase drift in OCT measurements follows an approximate 1/f behavior [25]. Variations are minimal over short timescales but can accumulate substantially over longer acquisitions. RevoPol acquires repeated B-scans over longer timescales ($\sim 10^{-2}$ s), making the reconstruction more susceptible to phase drift and motion artifacts. When no birefringent material is present along the light path, zero-phase correction informed by the surface signal can be applied, uniformly resetting the phase across all A-lines. In retinal imaging, however, corneal birefringence introduces cumulative retardance at the sample surface, invalidating this simple approach. In such cases, phase offset estimation is required. Since the cornea is considered a highly ordered birefringent tissue [43] and the surface retardance corresponds to a very small region, it is reasonable that we observed a uniform collagen structure with nearly constant retardance. Typically, the phase shift is assumed to be depth independent. However, we observed that adjacent B-scans exhibit abrupt phase changes within a single A-line in regions near blood vessels (**Fig. 8D**), which may introduce retardance artifacts. This observation violates the assumption of depth-independent phase shift. The phase shift inconsistency within single A-lines may arise from extravascular motion [44, 45]. To address this issue, our strategy is to estimate additional en face retardance and optic axis orientation maps at a specific depth using the phase optimization approach. After compensating for surface retardance, this approach enables more robust reconstruction of cumulative polarimetric properties at the specific depth.

Both surface retardance compensation and accurate phase correction rely on reliable surface segmentation. In retinal imaging, however, surface detection is often challenged by speckle noise, low signal contrast at certain boundaries, and local signal dropout caused by vessels or eye motion [46]. Conventional segmentation methods frequently fail to provide smooth and continuous delineation under these conditions, which in turn compromises the stability of the reconstructed polarimetric parameters. These limitations are not unique to RevoPol. To address this, we developed a $Q$ metric (Eq. 16) based on the ratio of the dominant singular value to the total energy of the coherency matrix. This metric, analogous to the degree of polarization, enhances surface contrast by suppressing speckle fluctuations while preserving structural boundaries. The $Q$ metric facilitates both surface segmentation and angiographic detection. The segmentation parameters were determined empirically and may require adjustment depending on image quality and tissue structure.

In summary, we demonstrated that RevoPol enables reliable reconstruction of retardance and optic axis orientation in both phantoms and human retina, requiring only minimal hardware modification to a commercial OCT platform. This framework extends the applicability of PS-OCT beyond retinal imaging and is suitable for both *ex vivo* and *in vivo* studies across diverse research and clinical settings. By lowering technical barriers, RevoPol facilitates broader adoption of polarization-sensitive OCT for diverse diagnostic, translational, and research applications.

## Acknowledgment

The authors gratefully acknowledge Drs. Ting-Yen Tsai and Ting-Hao Chen for their contributions to the early stages of this project. A pending U.S. patent application relates to the technology reported in this manuscript. No other conflicts of interest are declared.

## Data availability

All reconstruction algorithms, sample processing scripts, and test data will be openly available in the CBORT-NCBIB GitHub repository.

## References

[1] D. Huang *et al.*, "Optical coherence tomography," *Science,* vol. 254, no. 5035, pp. 1178-1181, 1991.

[2] W. Drexler and J. G. Fujimoto, "State-of-the-art retinal optical coherence tomography," *Progress in Retinal and Eye Research,* vol. 27, no. 1, pp. 45-88, 2008.

[3] E. A. Swanson and J. G. Fujimoto, "The ecosystem that powered the translation of OCT from fundamental research to clinical and commercial impact," *Biomedical Optics Express,* vol. 8, no. 3, pp. 1638-1664, 2017.

[4] G. J. Tearney *et al.*, "Consensus standards for acquisition, measurement, and reporting of intravascular optical coherence tomography studies: a report from the International Working Group for Intravascular Optical Coherence Tomography Standardization and Validation," *Journal of the American College of Cardiology,* vol. 59, no. 12, pp. 1058-1072, 2012.

[5] Z. Ali, K. K. Galougahi, G. S. Mintz, A. Maehara, R. Shlofmitz, and A. Mattesini, "Intracoronary optical coherence tomography: state of the art and future directions: State of the art in OCT," *EuroIntervention,* vol. 17, no. 2, p. e105, 2021.

[6] T.-H. Tsai *et al.*, "Optical coherence tomography in gastroenterology: a review and future outlook," *Journal of Biomedical Optics,* vol. 22, no. 12, pp. 121716-121716, 2017.

[7] H. Pahlevaninezhad and S. Lam, "Optical Coherence Tomography: A Review," *Interventions in Pulmonary Medicine,* pp. 379-391, 2023.

[8] B. Wan *et al.*, "Applications and future directions for optical coherence tomography in dermatology," *British Journal of Dermatology,* vol. 184, no. 6, pp. 1014-1022, 2021.

[9] M. R. Hee, D. Huang, E. A. Swanson, and J. G. Fujimoto, "Polarization-sensitive low-coherence reflectometer for birefringence characterization and ranging," *JOSA B,* vol. 9, no. 6, pp. 903-908, 1992.

[10] J. F. De Boer, C. K. Hitzenberger, and Y. Yasuno, "Polarization sensitive optical coherence tomography–a review," *Biomedical Optics Express,* vol. 8, no. 3, pp. 1838-1873, 2017.

[11] E. Li, S. Makita, Y.-J. Hong, D. Kasaragod, and Y. Yasuno, "Three-dimensional multi-contrast imaging of in vivo human skin by Jones matrix optical coherence tomography," *Biomedical Optics Express,* vol. 8, no. 3, pp. 1290-1305, 2017.

[12] K. H. Kim *et al.*, "In vivo imaging of human burn injuries with polarization-sensitive optical coherence tomography," *Journal of Biomedical Optics,* vol. 17, no. 6, pp. 066012-066012, 2012.

[13] W. C. Lo *et al.*, "Longitudinal, 3D imaging of collagen remodeling in murine hypertrophic scars in vivo using polarization-sensitive optical frequency domain imaging," *Journal of Investigative Dermatology,* vol. 136, no. 1, pp. 84-92, 2016.

[14] M. Villiger *et al.*, "Coronary plaque microstructure and composition modify optical polarization: a new endogenous contrast mechanism for optical frequency domain imaging," *JACC: Cardiovascular Imaging,* vol. 11, no. 11, pp. 1666-1676, 2018.

[15] K. Otsuka, M. Villiger, S. K. Nadkarni, and B. E. Bouma, "Intravascular polarimetry: clinical translation and future applications of catheter-based polarization sensitive optical frequency domain imaging," *Frontiers in Cardiovascular Medicine,* vol. 7, p. 146, 2020.

[16] D. C. Adams *et al.*, "Birefringence microscopy platform for assessing airway smooth muscle structure and function in vivo," *Science Translational Medicine,* vol. 8, no. 359, pp. 359ra131-359ra131, 2016.

[17] S. Nandy *et al.*, "Polarization-sensitive endobronchial optical coherence tomography for microscopic imaging of fibrosis in interstitial lung disease," *American Journal of Respiratory and Critical Care Medicine,* vol. 206, no. 7, pp. 905-910, 2022.

[18] R. R. Parakkel *et al.*, "Retinal nerve fiber layer damage assessment in glaucomatous eyes using retinal retardance measured by polarization-sensitive optical coherence tomography," *Translational Vision Science & Technology,* vol. 13, no. 5, pp. 9-9, 2024.

[19] S. Steiner *et al.*, "Birefringent properties of the peripapillary retinal nerve fiber layer in healthy and glaucoma subjects analyzed by polarization-sensitive OCT," *Investigative Ophthalmology & Visual Science,* vol. 63, no. 12, pp. 8-8, 2022.

[20] R. Terao *et al.*, "Quantification of Hyperreflective Foci in Age-Related Macular Degeneration by Polarization Sensitive Optical Coherence Tomography," *Ophthalmology Science,* p. 100792, 2025.

[21] X. Liu *et al.*, "Posterior scleral birefringence measured by triple-input polarization-sensitive imaging as a biomarker of myopia progression," *Nature Biomedical Engineering,* pp. 1-15, 2023.

[22] K. Ohno-Matsui *et al.*, "Polarization-sensitive OCT imaging of scleral abnormalities in eyes with high myopia and dome-shaped macula," *JAMA Ophthalmology,* vol. 142, no. 4, pp. 310-319, 2024.

[23] M. Villiger, B. Braaf, N. Lippok, K. Otsuka, S. K. Nadkarni, and B. E. Bouma, "Optic axis mapping with catheter-based polarization-sensitive optical coherence tomography," *Optica,* vol. 5, no. 10, pp. 1329-1337, 2018.

[24] N. Vansteenkiste, P. Vignolo, and A. Aspect, "Optical reversibility theorems for polarization: application to remote control of polarization," *Journal of the Optical Society of America A,* vol. 10, no. 10, pp. 2240-2245, 1993.

[25] B. J. Vakoc, S.-H. Yun, J. d. Boer, G. J. Tearney, and B. E. Bouma, "Phase-resolved optical frequency domain imaging," *Optics Express,* vol. 13, no. 14, pp. 5483-5493, 2005.

[26] X. Liu, K. Beaudette, X. Wang, L. Liu, B. E. Bouma, and M. Villiger, "Tissue-like phantoms for quantitative birefringence imaging," *Biomedical Optics Express,* vol. 8, no. 10, pp. 4454-4465, 2017.

[27] M. Pircher, E. Götzinger, B. Baumann, and C. K. Hitzenberger, "Corneal birefringence compensation for polarization sensitive optical coherence tomography of the human retina," *Journal of Biomedical Optics,* vol. 12, no. 4, pp. 041210-041210-10, 2007.

[28] B. Braaf, S. Donner, A. S. Nam, B. E. Bouma, and B. J. Vakoc, "Complex differential variance angiography with noise-bias correction for optical coherence tomography of the retina," *Biomedical Optics Express,* vol. 9, no. 2, pp. 486-506, 2018.

[29] M. Sugita *et al.*, "Motion artifact and speckle noise reduction in polarization sensitive optical coherence tomography by retinal tracking," *Biomedical Optics Express,* vol. 5, no. 1, pp. 106-122, 2013.

[30] B. Cense *et al.*, "Henle fiber layer phase retardation measured with polarization-sensitive optical coherence tomography," *Biomedical Optics Express,* vol. 4, no. 11, pp. 2296-2306, 2013.

[31] K. Twietmeyer, R. Chipman, A. E. Elsner, Y. Zhao, and D. VanNasdale, "Mueller matrix retinal imager with optimized polarization conditions," *Optics Express,* vol. 16, no. 26, pp. 21339-21354, 2008.

[32] N.-J. Jan *et al.*, "Polarization microscopy for characterizing fiber orientation of ocular tissues," *Biomedical Optics Express,* vol. 6, no. 12, pp. 4705-4718, 2015, doi: 10.1364/BOE.6.004705.

[33] C. Raoux, A. Chessel, P. Mahou, G. Latour, and M.-C. Schanne-Klein, "Unveiling the lamellar structure of the human cornea over its full thickness using polarization-resolved SHG microscopy," *Light: Science & Applications,* vol. 12, no. 1, p. 190, 2023.

[34] Y. Qu *et al.*, "Dichroism-sensitive photoacoustic computed tomography," *Optica,* vol. 5, no. 4, pp. 495-501, 2018.

[35] J. E. Roth, J. A. Kozak, S. Yazdanfar, A. M. Rollins, and J. A. Izatt, "Simplified method for polarization-sensitive optical coherence tomography," *Optics Letters,* vol. 26, no. 14, pp. 1069-1071, 2001.

[36] C. Fan and G. Yao, "Mapping local optical axis in birefringent samples using polarization-sensitive optical coherence tomography," *Journal of Biomedical Optics,* vol. 17, no. 11, pp. 110501-110501, 2012.

[37] C. Fan and G. Yao, "Mapping local retardance in birefringent samples using polarization sensitive optical coherence tomography," *Optics Letters,* vol. 37, no. 9, pp. 1415-1417, 2012.

[38] Q. Xiong *et al.*, "Constrained polarization evolution simplifies depth-resolved retardation measurements with polarization-sensitive optical coherence tomography," *Biomedical Optics Express,* vol. 10, no. 10, pp. 5207-5222, 2019.

[39] G. L. Jones, Q. Xiong, X. Liu, B. E. Bouma, and M. Villiger, "Single-input polarization-sensitive optical coherence tomography through a catheter," *Biomedical Optics Express,* vol. 14, no. 9, pp. 4609-4626, 2023.

[40] B. Baumann, W. Choi, B. Potsaid, D. Huang, J. S. Duker, and J. G. Fujimoto, "Swept source/Fourier domain polarization sensitive optical coherence tomography with a passive polarization delay unit," *Optics Express,* vol. 20, no. 9, pp. 10229-10241, 2012.

[41] Y. Lim, Y.-J. Hong, L. Duan, M. Yamanari, and Y. Yasuno, "Passive component based multifunctional Jones matrix swept source optical coherence tomography for Doppler and polarization imaging," *Optics Letters,* vol. 37, no. 11, pp. 1958-1960, 2012.

[42] B. H. Park *et al.*, "Real-time fiber-based multi-functional spectral-domain optical coherence tomography at 1.3 mu m," *Optics Express,* vol. 13, no. 11, pp. 3931-3944, 2005.

[43] K. M. Meek and C. Knupp, "Corneal structure and transparency," *Progress in Retinal and Eye Research,* vol. 49, pp. 1-16, 2015.

[44] A. Arrashoud, J. Reynaud, M. Dunn, S. K. Gardiner, G. Cull, and B. Fortune, "The Comb Artifact: Extravascular Motion Detected by OCT Angiography, a Novel Observation Related to Glaucoma," *Investigative Ophthalmology & Visual Science,* vol. 65, no. 7, pp. 5924-5924, 2024.

[45] A. Arrashoud, J. Reynaud, M. Dunn, G. Cull, S. K. Gardiner, and B. Fortune, "Extravascular Motion Signal Detected by OCT Angiography Indicates Altered Vascular-Tissue Biomechanical Interactions in Glaucoma," *Investigative Ophthalmology & Visual Science,* vol. 67, no. 4, pp. 10-10, 2026.

[46] M. A. Mayer, J. Hornegger, C. Y. Mardin, and R. P. Tornow, "Retinal nerve fiber layer segmentation on FD-OCT scans of normal subjects and glaucoma patients," *Biomedical Optics Express,* vol. 1, no. 5, pp. 1358-1383, 2010.